\documentclass[12pt]{article}

\usepackage[dvips]{graphicx}

\def\dspace{\baselineskip = 0.30in}

\def\lapproxeq{\lower .7ex\hbox{$\;\stackrel{\textstyle
<}{\sim}\;$}}
\def\gapproxeq{\lower .7ex\hbox{$\;\stackrel{\textstyle
>}{\sim}\;$}}

\begin{document}

\dspace

\begin{titlepage}
\begin{flushright}
BA-03-19  \\
\end{flushright}
\vskip 2cm
\begin{center}
{\Large\bf
Einstein Gravity on a Brane\\
in 5D Non-compact Flat Spacetime} \\
{\bf --DGP model revisited--}
\vskip 1cm {\normalsize\bf
Bumseok Kyae\footnote{bkyae@bartol.udel.edu} 
}
\vskip 0.5cm
{\it
Bartol Research Institute, University of Delaware, \\ 
Newark, DE~~19716,~~USA\\[0.1truecm]}

\end{center}
\vskip 1.5cm

\begin{abstract}

We revisit the 5D gravity model
by Dvali, Gabadadze, and Porrati (DGP).  
Within their framework it was shown that
even in 5D non-compact Minkowski space $(x^\mu,z)$,
the Newtonian gravity can emerge on a brane at short distances
by introducing a brane-localized 4D Einstein-Hilbert term 
$\delta(z)M_4^2\sqrt{|\bar{g}_4|}\bar{R}_4$ in the action. 
Based on this idea, 
we construct simple setups in which graviton standing waves can arise, 
and we introduce brane-localized $z$ derivative terms 
as a correction to $\delta(z)M_4^2\sqrt{|\bar{g}_4|}\bar{R}_4$.      
We show that the gravity potential of brane matter becomes $-\frac{1}{r}$
at {\it long} distances, because the brane-localized $z$ derivative terms allow
only a smooth graviton wave function near the brane.
Since the bulk gravity coupling may be arbitrarily small,
strongly interacting modes from the 5D graviton do not appear.
We note that the brane metric utilized to construct 
$\delta(z)M_4^2\sqrt{|\bar{g}_4|}\bar{R}_4$ can be relatively different 
from the bulk metric by a conformal factor,
and show that the graviton tensor structure 
that the 4D Einstein gravity predicts are reproduced in DGP type models. 

\end{abstract}
\end{titlepage}

\newpage



Since Kaluza and Klein proposed the five dimensional (5D) theories,
it had been believed for a long time that an extra space, if it exists,
should be compactified on an extremely small manifold.
The Newtonian gravity theory, which explains well
the observed gravity interactions,
seemingly ensures that our space should be effectively three dimensional.
As noted in Refs.~\cite{antoniadis,add,rs2},
however, the size of the extra dimension(s) could be
as large as ${\rm (TeV)^{-1}}$ scale~\cite{antoniadis,add},
and may even be infinite provided the graviton is effectively localized
on a four dimensional (4D) sub-space (brane) embedded
in a 5D AdS spacetime~\cite{rs2}.

Especially in Ref.~\cite{dgp},
Dvali, Gabadadze, and Porrati (DGP) argued
that the Newtonian gravity can be compatible
even with {\it 5D non-compact flat} spacetime,
only if (i) the relevant matter fields are localized on a 4D brane,
and (ii) a 4D Einstein-Hilbert term $M_4^2\sqrt{|\bar{g}_4|}\bar{R}_4$
is additionally introduced on the brane
apart from the bulk gravity kinetic term $M_5^3\sqrt{|g_5|}R_5$.
In Ref.~\cite{dgp}, it was claimed that
the ordinary Newtonian potential arises at short distances,
whereas at long distances the potential becomes that of a 5D theory.
Thus, $M_5$ should be supposed to be extremely small ($<<$ TeV)
so that the 4D gravity potential is modified
at longer distances than the Hubble length scale.
This setup was employed in the self-tuning model of
the cosmological constant~\cite{kks}.

As shown in Ref.~\cite{dgp}, however,
the graviton tensor structure in the 5D (minimal) DGP model is
given by that in tensor-scalar gravity theory
rather than that in the Einstein theory.\footnote{In Ref.~\cite{6ddgp},
it is demonstrated that in $D\geq 6$
the brane-localized gravity kinetic term exactly gives
the result of the Einstein gravity on the 4D brane.}
Thus, an extra scalar polarization degree 
is also involved in 4D gravity interaction.
This gives rise to unacceptable deviation
from the observation results on light bending around the sun\footnote{
To compensate the additional attractive force by the extra scalar mode,
the authors in Ref.~\cite{dgp} suggested to introduce a vector field,
which is universally coupled to all matter fields
with an $U(1)$ charge.  }
as in the massive gravity case~\cite{vDVZ}.\footnote{
In Ref.~\cite{vainshtein}, it was argued that
the resummation of nonlinear effects in massive gravity
recovers the result of the Einstein gravity near the sun.
This issue in DGP setup
is handled in Refs.~\cite{lightbending}. }
Moreover, in Ref.~\cite{lpr} the authors criticized the DGP model
pointing that extremely small $M_5$ possibly induces strong gravity 
interactions by $h_{\mu 5}$ and $h_{55}$
(whose kinetic terms were supposed to be provided only from 
$M_5^3\sqrt{|g_5|}R_5$ in the paper).
Hence, the validity of momentum expansion would break down
with extremely small $M_5$.

In this paper, we revisit the DGP model
with more considerable ingredients,
and discuss the long distance gravity potential
and the graviton tensor structure again.
We consider non-compact 5D spacetime $(x^\mu,z)$ ($\mu=0,1,2,3$) 
with the $Z_2$ symmetry, 
under which $z$ and $-z$ are identified.
We assign even (odd) parity of $Z_2$ to $g_{\mu\nu}$ and $g_{55}$
($g_{\mu 5}$).
Since 5D general covariance is explicitly broken
at the $Z_2$ fixed point (brane),
we require only 4D general covariance
on the brane~\cite{dgp}.
Let us consider the following action,
\begin{eqnarray} \label{action}
S=\int d^4xdz\bigg[\sqrt{|g_5|}\bigg(\frac{M_5^3}{2}R_5+{\cal L}_m^B\bigg)
+\delta(z)\sqrt{|\bar{g}_4|}\bigg(\frac{M_4^2}{2}\bar{R}_4
+{\cal L}_m^b \bigg)\bigg] ~,  
\end{eqnarray}
where $g_{5}\equiv {\rm Det}g_{MN}$ ($M,N=0,1,2,3,5$), 
$\bar{g}_{4}\equiv {\rm Det}g_{\mu\nu}$. 
${\cal L}_m^b$ (${\cal L}_m^B$) denotes brane (bulk) matter contributions 
to the action. In this paper, we regard all the standard model fields
as brane matter fields.   
In Eq.~(\ref{action}), we dropped the bulk cosmological constant 
and the brane tension.
We assume that they somehow vanish~\cite{kks}. 
While $R_5$ is the 5D Ricci scalar $g^{MN}R^P\,_{MPN}$,
$\bar{R}_4$ is defined as the 4D Ricci scalar
$\bar{g}^{\mu\nu}\bar{R}^\rho\,_{\mu\rho\nu}$ ($\mu,\nu,\rho=0,1,2,3$).
Even if $\bar{R}_4$ was not contained in the bare action,
it could be radiatively generated
below the conformal symmetry breaking scale~\cite{dgp}.  
%
Generically the metric $\bar{g}_{\mu\nu}$ defining $\bar{R}_4$
can be relatively different from the bulk metric $g_{MN}$
constructing $R_5$ by a scale factor $\omega^2(x,z)$,
\begin{eqnarray}
\bar{g}_{\mu\nu}(x,z)\equiv\delta_\mu^M\delta_\nu^Ng_{MN}(x,z)
\times\omega^2(x,z) ~,
\end{eqnarray}
which can not be removed by redefining the metric, and its degree
should appear in the bulk and/or on the brane.
$R_5$ and $\bar{R}_4$ constructed with
$g_{MN}$ and $\bar{g}_{\mu\nu}$ still respect
5D and 4D general covariance in the bulk and on the brane, respectively.
Since $\bar{R}_4$ will turn out to be dominant in gravity interaction
on the brane,  it is more convenient
to redefine the metric such that $\omega^2$ appears
only in the bulk side 
for proper interpretation of gravity interactions on the brane.
%

With vanishing bulk cosmological constant and brane tension,
the background metric should be flat, $\bar{g}_{\mu\nu}=\eta_{\mu\nu}$ and
$\omega^2$ a constant, which can be normalized to unity by rescaling $M_5$.
Beyond the leading term, however, $\omega^2$ would appear as a non-trivial
physical degree in the bulk.
The perturbed metric near the flat background is
\begin{eqnarray}
ds^2=
\bigg[1+\frac{1}{2}\phi(x,z)\bigg]^{-2}\bigg(\eta_{\mu\nu}+h_{\mu\nu}(x,z)
\bigg)dx^{\mu}dx^{\nu}+2h_{\mu 5}dx^\mu dz +\bigg(1+h_{55}(x,z)\bigg)dz^2 ~,
\end{eqnarray}
where $\eta_{\mu\nu}\equiv {\rm diag}(-1,1,1,1)$, and
$\phi$ indicates the sub-leading term of $\omega$,
i.e. $\omega^{-2}\approx (1+\frac{1}{2}\phi)^{-2}\approx (1-\phi)$.
On the other hand, on the brane the perturbed metric is just given
by $\eta_{\mu\nu}+h_{\mu\nu}$.

The localized gravity kinetic term $\delta(z)M_4^2\bar{R}_4$ 
in Eq.~(\ref{action}) adds a brane-localized 4D Einstein tensor 
to the 5D full gravity equation~\cite{dgp}.
At the linearized level, which is relevant in low energy gravity interactions,
it takes the form:
\begin{eqnarray} \label{localeinstein}
G_{\mu\nu}^{(0)}= -\frac{\delta(z)}{2M_5^3}M_4^2\bigg[
\nabla_4^2\bar{h}_{\mu\nu}
+\eta_{\mu\nu}\partial^{\lambda}\partial^{\delta}\bar{h}_{\lambda\delta}
-\partial_\mu\partial^\lambda\bar{h}_{\lambda\nu}
-\partial_\nu\partial^\lambda\bar{h}_{\lambda\mu}\bigg] ~,
\end{eqnarray}
where $\nabla_4^2$ denotes $\eta^{\mu\nu}\partial_\mu\partial_\nu$, and
we defined $\bar{h}_{\mu\nu}\equiv h_{\mu\nu}
-\frac{1}{2}\eta_{\mu\nu}h$, $h\equiv \eta^{\mu\nu}h_{\mu\nu}$.
The subscripts and superscripts are raised and lowered
with $\eta_{\mu\nu}$.
As is well known, the linearized Einstein tensor Eq.~(\ref{localeinstein})
is invariant under the gauge transformation,
$h_{\mu\nu}(x,z)\longrightarrow
h_{\mu\nu}(x,z)+\partial_\mu\xi_{\nu}(x,z)+\partial_\nu\xi_{\mu}(x,z)$.
Hence, it would be reasonable to consider also the following brane-localized 
higher derivative terms as a correction to Eq.~(\ref{localeinstein}),
\begin{eqnarray} \label{correction}
G_{\mu\nu}^{(1)}=-\alpha\frac{\delta(z)}{2M_5^3}\partial_z^2\bigg[
\nabla_4^2\bar{h}_{\mu\nu}
+\eta_{\mu\nu}\partial^{\lambda}\partial^{\delta}\bar{h}_{\lambda\delta}
-\partial_\mu\partial^\lambda\bar{h}_{\lambda\nu}
-\partial_\nu\partial^\lambda\bar{h}_{\lambda\mu}\bigg] ~,
\end{eqnarray}
where $\alpha$ is a dimensionless coupling and
$\partial_z^2\equiv \partial_z\partial_z$,
because Eq.~(\ref{correction}) still maintains the gauge symmetry and
the $Z_2$ symmetry.
Small brane excitation effects would appear as  
the correction by such $z$ derivative terms.
We note that the linearized tensor $G_{\mu\nu}^{(0)}+G_{\mu\nu}^{(1)}$
can be effectively obtained
by redefining $h_{\mu\nu}$ in Eq.~(\ref{localeinstein}) only on the brane
\begin{eqnarray}
h_{\mu\nu}\longrightarrow H_{\mu\nu}=
h_{\mu\nu}+\frac{\alpha}{M_4^2}\partial_z^2h_{\mu\nu} ~.
\end{eqnarray}

Since Eq.~(\ref{correction}) respects the gauge symmetry 
observed in Eq.~(\ref{localeinstein}), one could expect that 
it is somehow generated in higher energy scales.
To get Eqs.~(\ref{localeinstein}) and (\ref{correction})
in the equation of motion,
let us consider the following brane-localized gravity kinetic and
interaction terms in the linearized Lagrangian,
\begin{eqnarray} \label{linearaction}
{\cal L}_{\rm lin}=-\delta(z)\bigg[
\frac{M_4^2}{4}\bigg(
\frac{1}{2}(\partial_\mu H_{\nu\rho})^2
-\frac{1}{2}(\partial_\mu H)^2-(\partial^\nu H_{\mu\nu})^2
+\partial_\mu H\partial_\nu H^{\mu\nu}
\bigg)-\frac{1}{2}H_{\mu\nu}T^{\mu\nu}\bigg], 
\end{eqnarray}
where $H\equiv\eta^{\mu\nu}H_{\mu\nu}$
and $T_{\mu\nu}(x)$ indicates the energy-momentum tensor by brane-localized
matter fields.
Hence, unlike in the bulk metric, the perturbed metric on the brane is
effectively given by $H_{\mu\nu}$.
By embedding $\eta_{\mu\nu}+H_{\mu\nu}$ in the modified brane ``metric''
$\hat{g}_{\mu\nu}=\bar{g}_{\mu\nu}
+\frac{\alpha}{M_4^2}\partial_z^2\bar{g}_{\mu\nu}$, 
a generally covariant 4D Lagrangian can be constructed, 
$\delta(z)\sqrt{|\bar{g}_4|}~\bar{R}_4[\bar{g}_{\mu\nu}]
\longrightarrow\delta(z)\sqrt{|\hat{g}_4|}~\hat{R}_4[
\hat{g}_{\mu\nu}]$.  
A 4D general coordinate transformation still can be defined as 
$\partial_z^2\bar{g}'_{\mu\nu}
=\frac{\partial x^\rho}{\partial x^{'\mu}}
\frac{\partial x^\sigma}{\partial x^{'\nu}}\partial_z^2\bar{g}_{\rho\sigma}$
with $\partial_z^2(\frac{\partial x^\rho}{\partial x^{'\mu}})|_{z=0}=0$.  
The variation $\delta\partial_z^2 h_{\mu\nu}|_{z=0}$ of
Eq.~(\ref{linearaction}),
which is independent of $\delta h_{\mu\nu}|_{z=0}$,
leads to a constraint equation,
\begin{eqnarray} \label{constraint}
G_{\mu\nu}^{(0)}+G_{\mu\nu}^{(1)}=\frac{1}{M_5^3}\delta(z)T_{\mu\nu}(x)~.
\end{eqnarray}
Indeed, the variation $\delta\partial_z^2 h_{\mu\nu}|_{z=0}$
can not be converted to $\delta h_{\mu\nu}|_{z=0}$
through a partial integration, because the partial integration
for $\partial_z^2 h_{\mu\nu}|_{z=0}$ on the brane induces
physically ill-defined functions such as $\partial_z^2\delta(z)$
and $\partial_z\delta(z)$~\cite{sfv}.
The extremizing condition for ${\cal L}_{\rm lin}$
under $\delta h_{\mu\nu}$ yields the same expression,
but it takes part in the 5D gravity equation,
\begin{eqnarray} \label{extreme}
G^B_{\mu\nu}+G_{\mu\nu}^{(0)}+G_{\mu\nu}^{(1)}
=\frac{1}{M_5^3}\delta(z)T_{\mu\nu}(x)+\frac{1}{M_5^3}T^B_{\mu\nu}(x,z) ~, 
\end{eqnarray}
where $G^B_{\mu\nu}$ and $T^B_{\mu\nu}$ are the linearized bulk Einstein tensor 
and the bulk energy-momentum tensor, respectively.  
Hence, Eq.~(\ref{constraint}) implies  
$G_{\mu\nu}^{B}(x,z)=\frac{1}{M_5^3}T^B_{\mu\nu}(x,z)$.  
%
%

Once we introduce such brane-localized higher derivative terms  
shown in Eq.~(\ref{linearaction}),
in fact, it is perturbatively consistent
to consider also other higher order curvature terms like $R_5^2$,
$R_{MN}R^{MN}$, $R_{MNPQ}R^{MNPQ}$, etc.
In this paper, in order to see the effect by the terms 
in Eq.~(\ref{correction}) clearly in the linearized gravity equation, 
we assume that the higher order curvature terms in the Lagrangian 
are given by the Gauss-Bonnet type,
\begin{eqnarray}
\sqrt{|g_5|}\beta M_5\bigg(R_5^2-4R_{MN}^2+R_{MNPQ}^2
\bigg)+\delta(z)\sqrt{|\bar{g}_4|}\gamma\bigg(
\bar{R}_4^2-4\bar{R}_{\mu\nu}^2+
\bar{R}_{\mu\nu\rho\sigma}^2\bigg) ~, 
\end{eqnarray}
which leaves intact the linearized gravity equation 
derived from Eq.~(\ref{linearaction})
with the flat background spacetime~\cite{GB}.
Indeed, in supergravity the quadratic curvature terms
would appear as the Gauss-Bonnet type.
Moreover, even if supersymmetry is broken on the brane,
supersymmetry in the bulk can remain exact if the extra dimension size
is infinite~\cite{6ddgp,bulksusy}.
Actually, brane higher curvature terms with a combination different from
the Gauss-Bonnet ratio do not seriously change our conclusion.

With the 4D harmonic gauge,
$\partial^\mu (h_{\mu\nu}-\frac{1}{2}\eta_{\mu\nu}h)=0$, 
which fixes a gauge parameter $\xi_\mu(x,z)$, 
the linearized Einstein equation reads~\cite{kks,kkl}
\begin{eqnarray}
\label{mn}
&&(\mu\nu)~:~
\bigg[\partial_\mu\partial_\nu\bigg(\phi-\frac{1}{2}h_{55}\bigg)
-\eta_{\mu\nu}\nabla_4^2\bigg(\phi-\frac{1}{2}h_{55}\bigg)
-\frac{3}{2}\eta_{\mu\nu}\partial_z^2\phi\bigg] ~~~~~\nonumber \\
&&~~~~~~~~~+\frac{1}{2}\bigg[\partial_\mu\partial^zh_{\nu 5}
+\partial_\nu\partial^zh_{\mu 5}
-2\eta_{\mu\nu}\partial^\lambda\partial^zh_{\lambda 5}\bigg]
\nonumber \\
&&~~~~~~~~~-\frac{1}{2}\bigg[\nabla_5^2\bigg(h_{\mu\nu}
-\frac{1}{2}\eta_{\mu\nu}h\bigg)
-\frac{1}{2}\eta_{\mu\nu}\partial_z^2h\bigg]
\\
&&~~~~~-\frac{\delta(z)}{2M_5^3}\bigg[
\bigg(M_4^2+\alpha \partial_z^2\bigg)
\nabla_4^2\bigg(h_{\mu\nu}-\frac{1}{2}\eta_{\mu\nu}h\bigg)\bigg]
=\frac{1}{M_5^3}\delta(z)T_{\mu\nu}(x) ~,  \nonumber \\
\label{m5}
&&(\mu 5)~:~
-\frac{1}{2}\bigg[\nabla_4^2h_{\mu 5}
-\partial_\mu\partial^\lambda h_{\lambda 5}\bigg]
+\frac{3}{2}\partial_\mu\partial_z\phi
-\frac{1}{4}\partial_\mu\partial_zh =0 ~, \\
\label{55}
&&(55)~:~
-\frac{3}{2}\nabla_4^2\phi
+\frac{1}{4}\nabla_4^2h =0 ~,
\end{eqnarray}
where $\nabla_5^2$ ($\nabla_4^2$) indicates
$\eta^{\mu\nu}\partial_\mu\partial_\nu+\partial_z^2$ 
($\eta^{\mu\nu}\partial_\mu\partial_\nu$).
%
%
The terms in the fourth line of Eq.~(\ref{mn}) came from the brane-localized
terms in the action~Eq.~(\ref{action}).
%
The presence of non-vanishing but small energy-momentum tensor
by brane matter $T_{\mu\nu}$ are responsible for metric fluctuation near
the flat background.  
Here we neglected $T^B_{\mu\nu}(x,z)$ for simplicity.   
The equation from $\delta S/\delta\phi=0$ for the action $S$ turns out to be 
just $\eta^{MN}G_{MN}^B=\eta^{MN}T^B_{MN}=0$, 
which is consistent with Eq.~(\ref{constraint}).   

Eqs.~(\ref{mn}) and (\ref{m5}) are invariant under 
\begin{eqnarray}
&&h_{\mu 5}(x,z) \longrightarrow h_{\mu 5}(x,z)+\partial_\mu \xi_5(x,z) ~, \\
&&h_{55}(x,z) \longrightarrow h_{55}(x,z)+2\partial_z\xi_5(x,z) ~.  
\end{eqnarray}
We can choose $\xi_5(x,z)$ such that 
\begin{eqnarray}
h_{55}(x,z)=2\phi(x,z) 
\end{eqnarray}
is satisfied.  
From Eqs.~(\ref{mn}) and (\ref{m5}), the dynamics of $h_{\mu 5}$ is governed 
by $\partial_{(\mu }h_{\nu)5}=\eta_{\mu\nu}\partial^\lambda h_{\lambda 5}$ 
($=0$) and a boundary condition $h_{\mu 5}|_{z=0}=0$.  
%
%
Eqs.~(\ref{m5}) and (\ref{55}) are easily solved by setting
\begin{eqnarray} \label{F}
h(x,z)=6\phi(x,z) ~.
\end{eqnarray}
Then, the last terms of the first and third lines in Eq.~(\ref{mn})
cancel out, and so Eq.~(\ref{mn}) becomes much simpler    
\begin{eqnarray} \label{maineq}
\nabla_5^2\bigg(h_{\mu\nu}
-\frac{1}{2}\eta_{\mu\nu}h\bigg)
%
+\frac{\delta(z)}{M_5^3}\bigg[
\bigg(M_4^2+\alpha \partial_z^2\bigg)
\nabla_4^2\bigg(h_{\mu\nu}-\frac{1}{2}\eta_{\mu\nu}h\bigg)\bigg]
=-\frac{2\delta(z)}{M_5^3}T_{\mu\nu}(x) ~.  
\end{eqnarray}
Hence, the energy-momentum conservation law
$\partial^\mu T_{\mu\nu}=0$ is trivially satisfied.
After some algebra, Eq.~(\ref{maineq}) in 4D momentum space $(p,z)$ becomes 
\begin{eqnarray} \label{momeq}
\bigg[\bigg(-p^2+\partial_z^2\bigg)-\frac{\delta(z)}{M_5^3}\bigg(
M_4^2p^2+\alpha p^2\partial_z^2\bigg)\bigg]\tilde{h}_{\mu\nu}(p,z)
=-\frac{2\delta(z)}{M_5^3}\bigg[
\tilde{T}_{\mu\nu}(p)-\frac{1}{2}\eta_{\mu\nu}\tilde{T}(p)\bigg]~, 
\end{eqnarray}
where $p^2\equiv p^\mu p_\mu$ ($\leq 0$), 
$\tilde{T}\equiv \eta^{\mu\nu}\tilde{T}_{\mu\nu}$, 
and tildes indicates 4D Fourier-transformed fields.  
Note that the tensor structure of $\tilde{h}_{\mu\nu}$ is exactly
the same as that in the 4D Einstein gravity theory.
Hence, the scalar mode included in $h_{\mu\nu}$
is successively decoupled from low energy gravity interactions
between brane matter fields, and
only two degrees of freedom in polarization states of the graviton
survive as in the 4D Einstein gravity theory.
Actually, it was possible by considering $\omega^{-2}$ in the bulk metric.  

%
%
%
%
The bulk solution of Eq.~(\ref{momeq}) with the even parity 
would be given by a linear combination of ${\rm cos}kz$ and ${\rm sin}k|z|$,
where $k\equiv\sqrt{-p^\mu p_\mu}$ ($\geq 0$).
Their coefficients could be determined by the boundary condition at $z=0$.  
%
%
We note that $\partial_z^2({\rm sin}k|z|)$ generates a delta function.
Thus, from the last term in the left hand side of Eq.~(\ref{momeq}),
${\rm sin}k|z|$ induces a highly singular term 
proportional to $\delta^2(z)$, which
can not be matched to the right hand side of Eq.~(\ref{momeq}) 
in weakly coupled gravity theory.
This singularity can not be removed by introducing a suitable 
gravity counter term.  
Hence, the solution satisfying the boundary condition at $z=0$ 
should be given only by ${\rm cos}kz$.  
It explicitly satisfies also the constraint equation~(\ref{constraint}) 
(or $G^B_{\mu\nu}=0$).  

A ${\rm cos}kz$ type solution, however, implies that 
an outgoing wave ($e^{ik|z|}$) as well as an incoming wave ($e^{-ik|z|}$) 
should be generated when brane matter fields are fluctuating.  
This is inconsistent with causality.  
%
%
%
A simple way to naturally create incoming wave is to introduce two more branes 
around the $z=0$ brane.   
Then the right hand side of Eq.~(\ref{mn}) is modified into   
\begin{eqnarray} \label{mn'}
\frac{1}{M_5^3}\delta(z)T_{\mu\nu}(x)+\frac{1}{M_5^3}\bigg[\delta(z-z_c)
+\delta(z+z_c)\bigg]S_{\mu\nu}(x) ~, 
\end{eqnarray}
where the two additional branes are introduced 
symmetrically under $z\leftrightarrow -z$.  
This is a $T_{\mu\nu}-h_{\mu\nu}-S_{\mu\nu}$ coupled system.  
In this setup, the standing waves such as ${\rm cos}kz$, ${\rm sin}k|z|$
could arise between the $z=\pm z_c$ branes, 
while still only outgoing wave is allowed 
in the outside region of the two branes.   
But the term $\delta(z)\alpha k^2\partial_z^2h_{\mu\nu}$ in Eq.~(\ref{mn}) 
selects only ${\rm cos}kz$ type solution at $|z|\leq z_c$.    
Since low energy matter fluctuations would induce the graviton waves 
typically with long wave length, 
the additional branes can be located considerably far from the $z=0$ brane. 
%
%

The solutions satisfying such boundary conditions are  
\begin{eqnarray} \label{solI}
\tilde{h}^I_{\mu\nu}(k,z)&=&\frac{-2{\rm cos}kz}{M_4^2k^2-\alpha k^4}
\bigg[\tilde{T}_{\mu\nu}(k)-\frac{1}{2}\eta_{\mu\nu}\tilde{T}(k)\bigg] 
~~~~~~~~~~~~{\rm for}~~~|z|\leq z_c  ~, \\
\tilde{h}^{II}_{\mu\nu}(k,z)
%
%
&=&\frac{{\rm cos}kz_c~e^{ik|z|}}{iM_5^3k}\bigg[
\tilde{S}_{\mu\nu}(k)-\frac{1}{2}\eta_{\mu\nu}\tilde{S}(k)\bigg] 
~~~~~~~~~~~~~~{\rm for}~~~|z|\geq z_c~,  \label{solII}  
\end{eqnarray}
where $\tilde{S}_{\mu\nu}$ is determined such that 
the boundary condition at $|z|=z_c$ is fulfilled.    
$\tilde{S}_{\mu\nu}$ turns out to be related to $\tilde{T}_{\mu\nu}$,  
\begin{eqnarray}
\tilde{S}_{\mu\nu}(k)=\frac{-2M_5^3e^{-i(kz_c-\frac{\pi}{2})}}
{M_4^2k-\alpha k^3}\times \tilde{T}_{\mu\nu}(k) ~.  
\end{eqnarray}
Hence, gravity effects at the $z=0$ brane     
by ``dark matter'' fluctuations on the $z=\pm z_c$ branes 
would be very suppressed at low energy. 
Time evolution of $h_{\mu\nu}(x,z)$ is governed by  
$\tilde{T}_{\mu\nu}(k)$ ($=\int d^4x e^{ikx}T_{\mu\nu}(x)$).    

It is interesting to compare our solutions Eqs.~(\ref{solI}) and 
(\ref{solII}) with the solution in the original DGP model,   
\begin{eqnarray}
\sim \frac{e^{ik|z|}}{M_4^2k^2+2iM_5^3k} ~.  
\end{eqnarray}
In the DGP model, typical properties appearing in solutions of 5D theories 
($\sim \frac{1}{2iM_5^3k}$) and 4D theories ($\sim \frac{1}{M_4^2k^2}$) 
are contained in one solution. 
At low energies 5D property becomes dominant, 
while at high energies 4D property appears dominant.  
On the other hand, in our solution the two properties are separate 
as shown in Eqs. (\ref{solI}) and (\ref{solII}).   
In view of an observer living in the region $|z|>z_c$, 
$\tilde{T}_{\mu\nu}$ ($\tilde{S}_{\mu\nu}$) 
is negligible at low (high) energies.   
Since the observer at $|z|>z_c$ can not distinguish $\tilde{T}_{\mu\nu}$ and 
$\tilde{S}_{\mu\nu}$ if the relevant graviton's wave length is long enough 
$kz_c<<1$,  the resultant gravity effects are 
the same as those in the DGP model (upto the tensor structure) 
to the observer at $|z|>z_c$.        
However, to an observer living {\it in} the $z=0$ brane, 
the solution describing gravity interaction is always given 
by Eq.~(\ref{solI}).  

We note that at low energy 
Eq.~(\ref{solI}) guarantees the same gravity interaction 
on the $z=0$ brane as that in the 4D Einstein gravity theory. 
$M_5$ can be arbitrarily large, and so 
the strongly interacting modes from 5D graviton can be avoided.  
Only if $k^2<\frac{M_4^2}{\alpha}$, no ghost particle is excited 
in Eq.~(\ref{solI}).  
With $T_{00}(x)=\rho(x)>>T_{ii}(x)$ ($i=1,2,3$),
the non-relativistic low energy gravity potential on $z=0$ brane 
is calculated to be~\cite{dgp} 
\begin{eqnarray}
V(\vec{r})&=&\int dt\int\frac{d^4k}{(2\pi)^4}e^{-ikx}
\frac{1}{2}\tilde{h}_{00}(k,z=0) \nonumber \\
&\approx&-\frac{1}{8\pi M_4^2}\int d^3\vec{r}'
\frac{\rho(\vec{r}')}{|\vec{r}-\vec{r}'|} ~.  
\end{eqnarray} 
The Newtonian constant is determined to $G_N\equiv 1/(8\pi M_4^2)$.

In Eq.~(\ref{mn'}), we introduced two matter branes at $z=\pm z_c$ 
without any localized gravity kinetic terms.  
In fact, introduction of the $z$ derivative terms at $z=\pm z_c$ is 
dangerous, because they disallow outgoing waves 
also outside $z=\pm z_c$ branes.   
It is unnatural to introduce them  
only on the $z=0$ brane.  
%
%
[Thus, it would be more desirable to interpret the interval $|z|\leq z_c$ 
and $z=\pm z_c$ branes as the inside of a (thick) brane and its surfaces, 
respectively.]  

An alternative way to obtain a standing wave is to introduce bulk matter.  
Roughly speaking, introduction of bulk matter is nothing but to introduce 
infinite number of matter branes 
with making their interval lengths infinitely small. 
Again this setup is a $T_{\mu\nu}-h_{\mu\nu}-T^B_{\mu\nu}$   
coupled system, where $T^B_{\mu\nu}$ denotes the energy-momentum tensor 
contributed by bulk matter.  
In the presence of bulk matter, both outgoing and incoming waves are 
basically possible, and their mixing ratio in a solution would be determined 
by initial or boundary conditions. 
%
%
%
%
For simplicity, let us assume that a bulk matter field is distributed 
uniformly in the $z$ direction, i.e. $T^B_{MN}(x,z)=T^B_{MN}(x)$, and 
also assume $T^B_{\mu 5}=0$.  
This kind of energy momentum tensor can be provided 
by a bulk scalar field independent of $z$.      
As will be shown below, even when brane matter is absent,  
the bulk matter uniformly distributed in the $z$ direction compels 
outgoing and incoming graviton waves to be excited     
due to the presence of $-\frac{\delta(z)}{M_5^3}M_4^2p^2\tilde{h}_{\mu\nu}$ 
in Eq.~(\ref{momeq}).   

The right hand side of Eqs.~(\ref{mn}) and (\ref{55}) 
are modified into   
\begin{eqnarray} \label{modifiedmn}
&&(\mu\nu)~:~~\frac{1}{M_5^3}\delta(z)T_{\mu\nu}(x)
+\frac{1}{M_5^3}T^B_{\mu\nu}(x) ~, \\
&&(55)~:~~\frac{1}{M_5^3}T^B_{55}(x) ~.  
\end{eqnarray}
In that case, Eq.~(\ref{F}) should be replaced by 
\begin{eqnarray} \label{replacedeq}
\phi(x,z)=\frac{1}{6}\bigg[h(x,z)-f(x)\bigg] ~, 
\end{eqnarray}
where $f(x)$ satisfies 
$\frac{1}{4}\nabla_4^2f(x)=\frac{1}{M_5^3}T^B_{55}(x)$.  
For the modified equation of motion, 
\begin{eqnarray}
\bigg[\bigg(k^2+\partial_z^2\bigg)+\frac{\delta(z)}{M_5^3}\bigg(
M_4^2k^2+\alpha k^2\partial_z^2\bigg)\bigg]\tilde{h}_{\mu\nu}(k,z)
=-\frac{2}{M_5^3}\bigg[\delta(z)\tilde{J}_{\mu\nu}(k)+\tilde{J}^B_{\mu\nu}(k)
%
%
\bigg]~,    
\end{eqnarray}
where $\tilde{J}^{(B)}_{\mu\nu}\equiv\tilde{T}^{(B)}_{\mu\nu}
-\frac{1}{2}\eta_{\mu\nu}\tilde{T}^{(B)}$, 
we obtain the following solution, 
\begin{eqnarray} \label{solB}
\tilde{h}_{\mu\nu}(k,z)=-\frac{2\tilde{J}^B_{\mu\nu}(k)}{M_5^3k^2}
-\frac{2{\rm cos}kz}{M_4^2k^2-\alpha k^4}\bigg[\tilde{J}_{\mu\nu}(k)
-\frac{M_4^2}{M_5^3}\tilde{J}^B_{\mu\nu}(k)\bigg] ~.   
\end{eqnarray}
This solution is consistent with Eq.~(\ref{constraint}) 
(or $G_{\mu\nu}^{B}=\frac{1}{M_5^3}T^B(x,z)$).  
Were it not for the brane-localized kinetic terms, 
the last two terms with ${\rm cos}kz$ would just be a homogeneous part 
of the solution.  We note here that even if $\tilde{J}_{\mu\nu}=0$ 
but only if $\tilde{J}^B_{\mu\nu}\neq 0$,  
the graviton solution shows the $z$ dependence.  
%
%
When $\tilde{J}_{\mu\nu}\neq 0$ and $\tilde{J}^B_{\mu\nu}\neq 0$, 
bulk matter enables the graviton wave to satisfy the boundary conditions  
by enhancing its incoming part.    
Since the low energy approximate solution at $z=0$ is    
\begin{eqnarray} \label{sol0}
\tilde{h}_{\mu\nu}(k,z=0)\approx 
-\frac{2\tilde{J}_{\mu\nu}(k)}{M_4^2k^2}\bigg[1+
O\bigg(\frac{\alpha k^2}{M_4^2}\bigg))\bigg] ~, 
%
\end{eqnarray}
the leading term of the graviton solution at low energy 
coincides with the solution in the 4D Einstein gravity theory.    

Although we showed that the 4D Einstein gravity is reproduced on the brane 
with a special bulk matter field, we arrive at the same conclusion 
also with a general $T^B_{MN}(x,z)$ by bulk matter.   
In that case $f(x)$ in Eq.~(\ref{replacedeq}) should be generalized to 
$f(x,z)$.  
For any bulk source $S^B_{\mu\nu}(x,z)$ contributed by $T^B_{\mu\nu}(x,z)$,  
$\frac{1}{4}\eta_{\mu\nu}\partial_z^2f(x,z)$, 
$\frac{1}{2}\partial_\mu\partial^zh_{\nu 5}+\cdots$, and so on, 
the bulk solution generally has the following form, 
\begin{eqnarray} 
\tilde{h}_{\mu\nu}(k,z)=-\frac{2}{M_5^3}\int\frac{dk_5}{2\pi}
\frac{{\rm cos}k_5z}{k^2-k_5^2}\tilde{S}^B_{\mu\nu}(k,k_5)
+p_{\mu\nu}(k){\rm cos}kz ~,
\end{eqnarray}
where $p_{\mu\nu}(k)$ should be determined by the boundary condition at $z=0$.  
Because of the boundary condition by $\delta(z)M_4^2k^2\tilde{h}_{\mu\nu}$, 
the solution at $z=0$ reduces to Eq.~(\ref{sol0}) upto $O(\alpha k^2/M_4^2)$   
for any arbitrary $S^B_{\mu\nu}(x,z)$.   
%
%
%
%
%
%

In conclusion, we have shown that the 5D DGP type gravitational models can be
phenomenologically viable.     
We introduced two more branes and/or bulk matter to make the graviton's
standing waves possible.  
Apart from bulk and brane gravity kinetic terms $M_5^3\sqrt{|g_5|}R_5$,    
$\delta(z)M_4^2\sqrt{|\bar{g}_4|}\bar{R}_4$ as in the original DGP model,
we consider also the brane-localized $z$ derivative terms    
at the linearized level 
in order to allow only smooth graviton waves near the brane.   
In this model, the long distance gravity potential on the brane
turns out to be the Newtonian potential.    
Since $M_5$ can be arbitrarily large,
the strongly interacting modes from the 5D graviton can be avoided. 
Since the brane metric can be relatively different from the bulk metric
by a conformal factor, we can obtain the desired tensor structure 
of the graviton.   
%
%

\vskip 0.3cm

\noindent{\bf Acknowledgments}

\noindent
The work is partially supported
by DOE under contract number DE-FG02-91ER40626.  
%


\begin{thebibliography}{99}

\bibitem{antoniadis} I.~Antoniadis,
Phys.\ Lett.\ B {\bf 246} (1990) 377. 

\bibitem{add} 
N.~Arkani-Hamed, S.~Dimopoulos and G.~R.~Dvali,
Phys.\ Lett.\ B {\bf 429} (1998) 263
[arXiv:hep-ph/9803315];
I.~Antoniadis, N.~Arkani-Hamed, S.~Dimopoulos and G.~R.~Dvali,
Phys.\ Lett.\ B {\bf 436} (1998) 257
[arXiv:hep-ph/9804398].

\bibitem{rs2} L.~Randall and R.~Sundrum,
Phys.\ Rev.\ Lett.\  {\bf 83} (1999) 4690
[arXiv:hep-th/9906064].

\bibitem{dgp} G.~R.~Dvali, G.~Gabadadze and M.~Porrati,
Phys.\ Lett.\ B {\bf 485} (2000) 208
[arXiv:hep-th/0005016].

\bibitem{kks} J.~E.~Kim, B.~Kyae and Q.~Shafi,
arXiv:hep-th/0305239.

%
\bibitem{6ddgp} G.~R.~Dvali and G.~Gabadadze,
Phys.\ Rev.\ D {\bf 63} (2001) 065007
[arXiv:hep-th/0008054].

\bibitem{vDVZ} H.~van Dam and M.~J.~Veltman,
Nucl.\ Phys.\ B {\bf 22} (1970) 397;
V.~I.~Zakharov, JEPT\ Lett. {\bf 12} (1970) 312.

\bibitem{vainshtein} A.~I.~Vainshtein,
Phys.\ Lett.\ B {\bf 39} (1972) 393.
See also T.~Damour, I.~I.~Kogan and A.~Papazoglou,
Phys.\ Rev.\ D {\bf 67} (2003) 064009
[arXiv:hep-th/0212155].

\bibitem{lightbending}
M.~Porrati,
Phys.\ Lett.\ B {\bf 534} (2002) 209
[arXiv:hep-th/0203014];
T.~Tanaka,
arXiv:gr-qc/0305031; 
C.~Middleton and G.~Siopsis,
arXiv:hep-th/0311070.

\bibitem{lpr} M.~A.~Luty, M.~Porrati and R.~Rattazzi,
JHEP {\bf 0309} (2003) 029
[arXiv:hep-th/0303116].
See also 
V.~A.~Rubakov,
arXiv:hep-th/0303125.

\bibitem{sfv} F.~del Aguila, M.~Perez-Victoria and J.~Santiago,
JHEP {\bf 0302} (2003) 051
[arXiv:hep-th/0302023].

\bibitem{GB} J.~E.~Kim, B.~Kyae and H.~M.~Lee,
Nucl.\ Phys.\ B {\bf 582} (2000) 296
[Erratum-ibid.\ B {\bf 591} (2000) 587]
[arXiv:hep-th/0004005];
J.~E.~Kim and H.~M.~Lee,
Nucl.\ Phys.\ B {\bf 602} (2001) 346
[Erratum-ibid.\ B {\bf 619} (2001) 763]
[arXiv:hep-th/0010093];
J.~E.~Kim, B.~Kyae and H.~M.~Lee,
Phys.\ Rev.\ D {\bf 62} (2000) 045013
[arXiv:hep-ph/9912344].
%

\bibitem{bulksusy} 
G.~R.~Dvali and M.~A.~Shifman,
Nucl.\ Phys.\ B {\bf 504} (1997) 127
[arXiv:hep-th/9611213]. 

\bibitem{kkl} J.~E.~Kim, B.~Kyae and H.~M.~Lee,
Phys.\ Rev.\ D {\bf 66} (2002) 106004
[arXiv:hep-th/0110103].

%
%



\end{thebibliography}
\end{document}